\documentclass[10pt,twocolumn,letterpaper]{article}

\usepackage{cvpr}              

\definecolor{cvprblue}{rgb}{0.21,0.49,0.74}
\usepackage[pagebackref,breaklinks,colorlinks,allcolors=cvprblue]{hyperref}

\usepackage{url}
\usepackage{amsmath}
\usepackage{amssymb}
\usepackage{mathtools}
\usepackage{amsthm}
\usepackage{multirow,tabularx}
\usepackage{longtable}
\usepackage{booktabs,longtable}
\usepackage{hhline}
\usepackage{microtype}
\usepackage{graphicx}
\usepackage{booktabs}
\usepackage{algorithmic}
\usepackage{algorithm}
\usepackage{pifont}
\usepackage{enumitem}
\usepackage{tcolorbox}
\usepackage{wrapfig}
\usepackage{amsmath}
\usepackage{bm}
\usepackage{colortbl} 
\usepackage[accsupp]{axessibility}  

\definecolor{mgray}{RGB}{240, 240, 240}
\def\paperID{8600} 
\def\confName{CVPR}
\def\confYear{2025}

\title{Enhancing Dance-to-Music Generation via Negative Conditioning Latent Diffusion Model} 

\author{Changchang Sun$^{1}$ ~~~~Gaowen Liu$^{2}$ ~~~~Charles Fleming$^2$  ~~~~Yan Yan$^1$\thanks{corresponding author}\\
  $^1$University of Illinois Chicago
  ~~~~~ $^2$Cisco Research
  \\
  {\tt\small \{csun47, yyan55\}@uic.edu}
  ~~~~~ {\tt\small \{gaoliu, chflemin\}@cisco.com}
}

\begin{document}
\maketitle
\begin{abstract}
Conditional diffusion models have gained increasing attention since their impressive results for cross-modal synthesis, where the strong alignment between conditioning input and generated output can be achieved by training a time-conditioned U-Net augmented with cross-attention mechanism. In this paper, we focus on the problem of generating music synchronized with rhythmic visual cues of the given dance video. Considering that bi-directional guidance is more beneficial for training a diffusion model, we propose to enhance the quality of generated music and its synchronization with dance videos by adopting both positive rhythmic information and negative ones (PN-Diffusion) as conditions, where a dual diffusion and reverse processes is devised. Specifically, to train a sequential multi-modal U-Net structure, PN-Diffusion consists of a noise prediction objective for positive conditioning and an additional noise prediction objective for negative conditioning. To accurately define and select both positive and negative conditioning, we ingeniously utilize temporal correlations in dance videos, capturing positive and negative rhythmic cues by playing them forward and backward, respectively. Through subjective and objective evaluations of input-output correspondence in terms of dance-music beat alignment and the quality of generated music, experimental results on the AIST++ and TikTok dance video datasets demonstrate that our model outperforms SOTA dance-to-music generation models.
\end{abstract}    
\section{Introduction}
\label{sec:intro}
Recent years have witnessed the unprecedented development of AI-powered realistic content generation in image, video, and audio domains,
thanks to the impressive synthesis capabilities of diffusion models.
From a task-level perspective, existing diffusion models can be roughly classified into two lines: unconditional diffusion models~\cite{abs-2207-12598,DhariwalN21} and conditional diffusion models~\cite{NicholDRSMMSC22,abs-2204-06125,RombachBLEO22,SahariaCSLWDGLA22}.
Specifically, unconditional diffusion models generate data by utilizing noise sampled from Gaussian distributions as input, while conditional ones typically incorporate noise along with embedding feature condition extracted from some specific modalities.
Due to the growing demand of customization and the flourish of multimodel data, increasing efforts have been dedicated to the conditional manner, giving rise to the emerging real-world application of cross-model generation, \textit{e.g.}, Text-to-Image (T2I) generation~\cite{RombachBLEO22,abs-2210-09549,SahariaCSLWDGLA22,abs-2204-06125}, Text-to-Audio (T2A) generation~\cite{HuangHY0LLYLYZ23,abs-2302-03917,LiuCYMLM0P23}, and Text-to-Video (T2V) generation~\cite{abs-2310-08465,abs-2305-13840,ho2022video}.
In this paper, we focus on the dance-to-music (D2M) generation task and target at 
training a latent diffusion model (LDM),
which has significant practical application in video-sharing platforms such as TikTok\footnote{\url{https://www.tiktok.com/.}} and YouTube\footnote{\url{https://www.youtube.com/.}} to generate appropriate music accompaniments for dance videos uploaded by users. 

 \begin{figure}[t]
	\centering
	\includegraphics[scale=0.147]{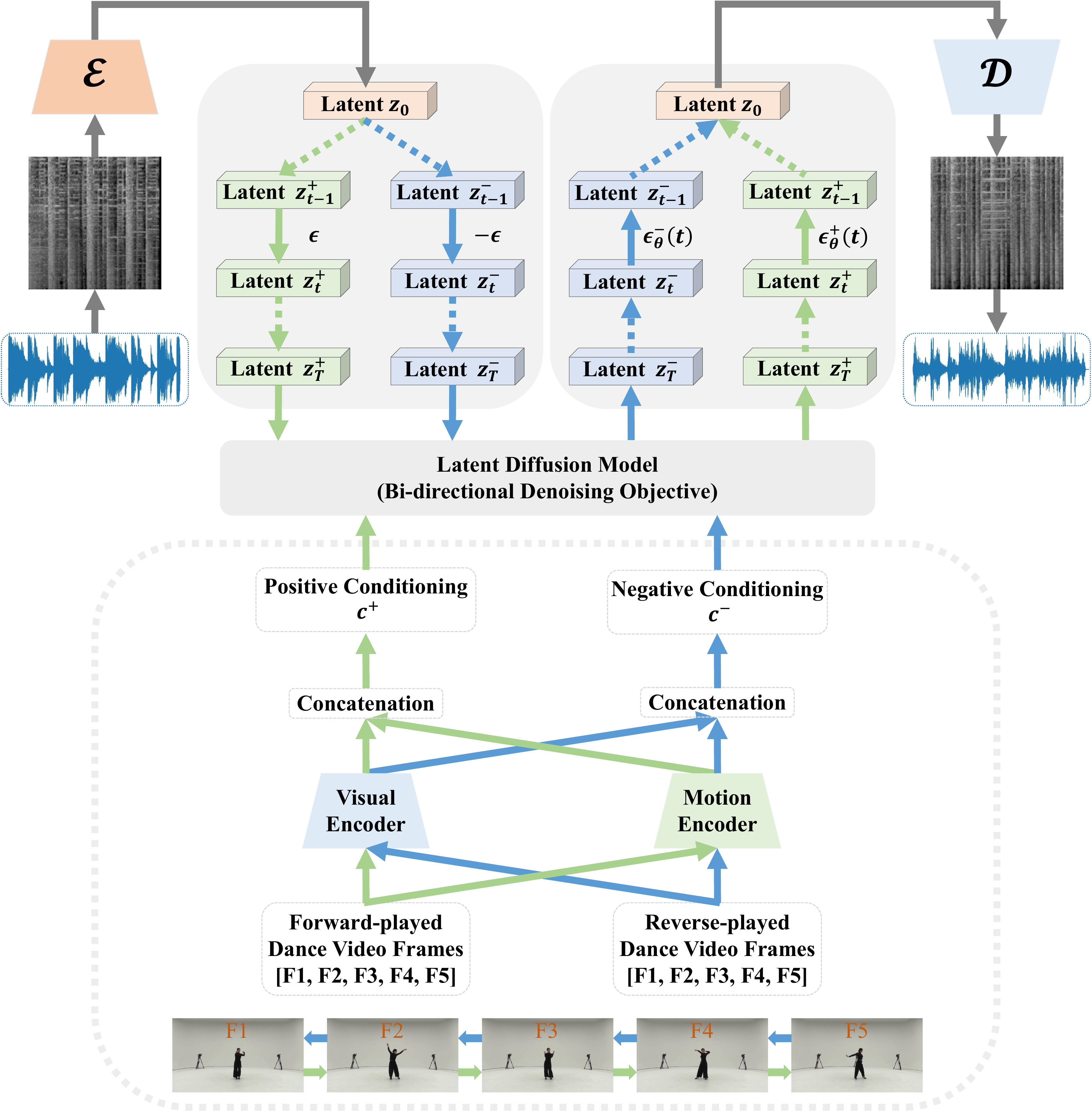}
\vspace{-0.1in}
	\caption{
 Illustration of our proposed PN-Diffusion. 
 The rich temporal synchronization information conveyed by the normal forward-played dance videos and their reverse-played counterparts are extracted by visual and motion encoders.
 A dual diffusion process and a dual reserve process are introduced to better realize the temporal correlation and rhythmic consistency between dance video and music, and a bi-directional denoising objective is designed to train the diffusion model.
}
\label{framework}
\vspace{-0.3in}
\end{figure}

To generate music accompaniments whose rhythms are coordinated and harmonious to the dance movements,
understanding the temporal correlation and rhythmic consistency between dance video and music is crucial, posing greater challenges compared to other conditional music generation tasks, such as text-to-music generation~\cite{abs-2302-03917}.
Generally, existing diffusion-based D2M methods incline to first 
extract visual rhythm and motion information from the dance video and then embed these multiple conditions into the input layer of the U-Net network, so as to guide and influence the training and inference of the diffusion model.
For example, CDCD~\cite{Zhu0ORT023} and LORIS~\cite{YuWCS023} put forward conditional diffusion models where the human body motions of
dance performers and the visual representation obtained from dance video frames are concatenated to form the final continuous conditioning input.

However, 
existing diffusion-based D2M generation works~\cite{Zhu0ORT023,YuWCS023}
fail to fully explore the special properties of dance video, where only the positive rhythmic visual cues and motion information are extracted from the dance video played in the normal forward direction to establish the temporal correlation
and rhythmic consistency, overlooking the potential effect of negative rhythm information conveyed by reverse-played dance video.
In fact, similar to human learning and machine learning where
positive and negative examples are essential to provide bi-directional guidance,
normal forward-played dance videos and their reverse-played counterparts can both offer distinct rhythms and motion information.
Positive examples guide the model on what to do, highlighting the desirable features that should be retained, while negative examples
serve as cautionary instances and instruct the model on what to avoid, pointing out undesirable characteristics.
Therefore,
it is favorable to introduce the negative rhythmic visual cues and negative motion information in the training process of the D2M diffusion models, realizing better temporal correlation and rhythmic
consistency between dancer movement and music.

The main challenge in incorporating negative conditioning into training a conditional diffusion probabilistic model lies in 
seamlessly 
integrating it with prior positive conditioning within the sequential multi-modal U-Net architecture.
Taking the noise prediction objective of diffusion model for example,
whether to have the negative conditioning directly influence the prediction of the noise added in the forward process or to introduce a negative noise during the forward process is another aspect that needs to be carefully considered.
Towards this end,
we propose PN-Diffusion, a negative conditioning latent diffusion model,
to
enhance the D2M synthesis, as shown in Fig.~\ref{framework}.
On the one hand,
we simultaneously exploit the rich temporal synchronization information 
conveyed by the normal forward-played dance videos
and their reverse-played counterparts, 
termed as positive conditioning and negative conditioning, respectively.
On the other hand, a dual diffusion and reverse processes are designed in our PN-Diffusion model.
Specifically,
we concurrently introduce the positive diffusion process and the negative diffusion process, whose noises are added separately with identical and opposite numerical values in each forward process step. 
Meanwhile, corresponding positive reverse process and negative reverse process are devised, where positive conditioning is adopted to guide the prediction of noise added in the positive diffusion process and negative conditioning is utilized to influence the prediction of the noise added in the negative diffusion process.
In this way, 
the performance of 
the sequential multi-modal U-Net architecture can be improved 
by simultaneously exploiting the rich positive and negative temporal synchronization
conveyed by the dance video played in normal and reverse direction.
To verify the effectiveness of proposed PN-Diffusion, extensive experiments are conducted on two 
real-world
dance video 
datasets AIST++~\cite{LiYRK21} and TikTok~\cite{ZhuOWACYT22}.
Our main contributions can be summarized in threefold:
\begin{itemize}[leftmargin=*]
	\item 
 To the best of our knowledge, we are the first attempt to 
 take advantage of the positive conditioning and negative conditioning simultaneously to enhance the performance of diffusion-based dance-to-music generation model.
 \item 
 A dual diffusion and reverse processes are devised in our PN-Diffusion model, and a new noise prediction objective 
 involving the noise prediction in the positive reverse process and negative reverse process is introduced. 
	\item 
 Extensive experiments demonstrate the superiority of our model 
regarding the correspondence of dance-music beats and the quality of generated music on AIST++\footnote{\url{https://youtu.be/FIbimVCnbWY}} and Tiktok\footnote{\url{https://youtube.com/shorts/IlapZ_Uj-Yg?feature=share}} datasets.
Codes are available at \url{https://github.com/Changchangsun/PN-Diffusion}.
\end{itemize}

\section{Related Work}

\subsection{Uni-modal Music Generation}
Uni-modal Music Generation~\cite{DongHYY18,EngelACGDR19,OordDZSVGKSK16,abs-2111-05011,dhariwal2020jukebox} focuses on generating editable music
based on the pre-established music representations.
One group of methods that has been extensively explored for audio modelling are Generative Adversarial Networks (GANs)~\cite{goodfellow2020generative}.
For example, MuseGAN~\cite{DongHYY18} and GANSYNTH~\cite{EngelACGDR19} introduce the GAN-based model that 
can generate multi-track and high-fidelity music.
Apart from GAN-based models, to generate novel and highly realistic musical fragments, WaveNet~\cite{OordDZSVGKSK16} presents a deep neural network that is
fully probabilistic and autoregressive, where the predictive distribution for each audio sample is conditioned on all preceding samples.
Rave~\cite{abs-2111-05011} designs a real-time audio variational autoencoder (VAE), allowing both fast and high-quality audio waveform synthesis.
In addition, to compress raw audio to a lower-dimensional discrete space, Jukebox~\cite{dhariwal2020jukebox} put forwards a multiscale
Vector Quantized Variational Autoencoders (VQ-VAE) to compress long context of raw audio to discrete codes.
Although existing uni-model music generation methods have achieved significant progress, they can not meet the real-world demands of generating specific types and contents with human control.

\subsection{Cross-Modal Music Generation}
Recently, researchers have delved into cross-modality music generation, including text-to-music~\cite{huang2023noise2music,agostinelli2023musiclm} and visual-to-music~\cite{su2021does,Zhu0ORT023,YuWCS023}, which works on constructing inter-modality correlation as supervision signals to enhance the quality and diversity of generated music.
Regarding text-to-music generation, 
Noise2Music~\cite{huang2023noise2music}
formulates a series of diffusion models to generate high-quality
$30$-second music clips from text prompts. 
As for visual-to-music generation, 
D2M-GAN~\cite{ZhuOWACYT22} presents an adversarial multi-modal framework to generate complex musical samples conditioned on dance video frames and human body motions.
CMT~\cite{DiJ0WZHLY21} devises a controllable music transformer to generate background music that matches the
given video.
Taking advantage of diffusion probabilistic models (DPMs),
CDCD~\cite{Zhu0ORT023} and LORIS~\cite{YuWCS023} adopt a latent diffusion probabilistic model to perform conditional audio generation.

\subsection{Latent Diffusion Model} 
Diffusion Probabilistic Models (DPMs)~\cite{abs-2207-12598,DhariwalN21} represent a novel class of likelihood-based generative models that have demonstrated remarkable performance to  image~\cite{abs-2210-09549,abs-2204-06125} and audio synthesis tasks~\cite{Zhu0ORT023,abs-2302-03917}, decomposing the data formation process into a sequential application of denoising autoencoders.
These models involve a dual process: a forward process that  
 slowly adds random noise to signal and a reverse process that gradually 
 constructs desired data samples from the noise. 
In contrast to Variational Autoencoders (VAEs)~\cite{doersch2016tutorial} or flow models~\cite{doersch2016tutorial}, diffusion models are trained using a predetermined procedure and the latent variable maintains high dimensionality identical to that of the original data.
The optimization of powerful DPMs often consumes hundreds of GPU
days and the inference is expensive since sequential evaluations.
In order to make DPMs more
practical, Latent Diffusion Model(LDM)~\cite{RombachBLEO22} applies DPMs training in the latent space of powerful pre-trained autoencoders, 
thereby reducing the number of pixels and accelerating the sampling speed. 
In our work, we also transform the original music sampled at specific sample rate into the Mel-spectrograms, and train an latent diffusion model.

\section{Methodology}
This section presents PN-Diffusion, our latent diffusion model with negative conditioning for realistic music generation from dance videos. We first introduce the music spectrogram and conditioning (Sec.\ref{3.1}), then detail the dual diffusion and reverse processes of PN-Diffusion (Sec.\ref{3.2}). Finally, we describe our U-Net architecture with positive and negative conditioning and derive the objective function (Sec.~\ref{3.3}).

\sloppy

\subsection{Music Spectrogram and Conditioning}\label{3.1}
\noindent\textbf{Music Spectrogram.}
For the music part, the input music audio is converted to the Mel spectrogram and  a Variational Autoencoder (VAE) is trained to learn the probabilistic mapping between input data and latent spectrogram space. 
Suppose that we have a set of $N$ dance videos $\mathcal{V}=\left\lbrace m_{i}, d_{i}\right\rbrace_{i=1}^{N}$.
For the raw musical audios, 
we set the sample rate as $22,050$ $\mathrm{Hz}$ and the Mel Basis of Mel-spectrograms M equals to $256$. 
In this way, we obtain a set of 
audio spectrogram $\mathcal{A}=\left\lbrace \mathbf{a}_{i}\right\rbrace_{i=1}^{N}$, whose dimension is $256\times256$.


\noindent\textbf{Perceptual Image Compression.}
Following LDM~\cite{RombachBLEO22}, our perceptual image compression model also consists of an encoder $\mathcal{E}$ and decoder $\mathcal{D}$, which is trained by combination of a perceptual loss~\cite{zhang2018unreasonable} and a patch-based adversarial objective~\cite{yu2021vector}.
Formally, given the audio spectrogram $\mathbf{a}_i\in\left[0,1\right]^{256\times256}$, the encoder $\mathcal{E}$ encodes $\mathbf{a}_i$ into a latent representation $\mathbf{z}_i=\mathcal{E}(\mathbf{a}_i)$, and the decoder $\mathcal{D}$ reconstructs the image from the spectrogram latent space $\tilde{\mathbf{a}_i}=\mathcal{D}(\mathbf{z}_i)$, where $\mathbf{z}_i\in \mathbb{R}^{32\times32}$. Then,  
the Gaussian noise is added and removed to/from the compressed latent spectrogram space using a linear noise schedule, and 
our PN-Diffusion model is trained on this latent spectrogram space conditioned on the visual cues and motion information.


\noindent\textbf{Positive and Negative Conditioning.}
In our work, we adopt two types of feature embeddings extracted from the dance video frames as the condition, named as rhythmic visual cues and motion information.
Specifically, for the rhythmic visual cues,
following previous waveform-based method~\cite{Zhu0ORT023},
we adopt the I3D (Inflated 3D ConvNet~\cite{yadav2022inflated}) model to extract the visual embeddings $p_i\in\mathbb{R}^{2048}$ from the video frames sampled from dance video $d_i$. Besides, we extract the 2D skeletons of dancer in each frame of the video using BlazePose~\cite{abs-2206-11678}, 
which can be organized into a graph sequence according to the direction in which the video frame plays.
And then we design a motion encoder based on the spatial temporal graph convolutional networks
ST\_GCN~\cite{YanXL18} to obtain the motion information $\mathbf{q}_i\in\mathbb{R}^{1024}$. Thereafter, these two kinds of features are concatenate together to act as conditioning $\mathbf{c}$. 

Considering that bi-directional guidance is more beneficial for training a diffusion model, 
we adopt both positive rhythmic information and negative ones as conditions to train the U-Net.
However, how to define and select the accurate negative conditioning to enhance the generation power of diffusion model constitutes a tough challenge.
In a sense, the higher the negativity of negative prompt, the more significant their adverse impact during the diffusion training process. 
Regarding our D2M task, 
it is difficult to directly provide negative samples that are entirely opposite to the given positive dance video samples, and
there is also no existing work to define and obtain the negative conditioning from a given dance video.
Towards this end, we target at fully exploring the special properties of dance video, primarily involving changes in temporal and motion. 
Specifically, we assume that forward-played and their reverse-played dance videos provide
distinct rhythms and motion information.
In a sense, reverse-played videos preserve the same poses, transitions, and temporal structure as forward-played ones but in the opposite direction, forming a more faithful negative pairing.
We can obtain positive conditioning $\mathbf{c}^{+}$ and negative conditioning $\mathbf{c}^{-}$ by playing the dance video in both forward and reverse order,
respectively.


\subsection{Dual Diffusion and Reverse Processes}\label{3.2}
Diffusion models are powerful generative models
that first transfer a given data distribution $z_0$ into unstructured noise (Gaussian noise in practice) in the forward diffusion process,
by gradually adding Gaussian noise $\epsilon\sim\mathcal{N}(0,I)$ 
to the data according to a variance schedule $\{\beta_1,\cdots,\beta_T\}$. This process can be expressed as follows,
\vspace{-0.05in}
\begin{equation}
\begin{aligned}
&q(z_t|z_{t-1})= \mathcal{N}(z_t;\sqrt{1-\beta_{t}}z_{t-1},\beta_{t}I),\\
&q(z_{1:T}|z_{0})= \prod_{t=1}^{T}q(z_t|z_{t-1}).
\end{aligned}
\vspace{-0.05in}
\end{equation}
Then, to recover the original data $z_0$ from a probability density $p(z_T)$, the random noise is iteratively denoised
through a fixed Markov Chain of length $T$
by a sequence of denoising autoencoders $\theta$ in the reverse process.
That is,
\vspace{-0.05in}
\begin{equation}
\begin{aligned}
&p(z_{t-1}|z_{t})= \mathcal{N}(z_{t-1};\mu_{\theta}(z_t,t),\Sigma_\theta(z_{t},t)),\\
&p(z_{0:T})= p(z_T)\prod_{t=1}^{T}p_\theta(z_{t-1}|z_{t}),
\end{aligned}
\vspace{-0.05in}
\end{equation}
where $\mu_{\theta}$ denotes the Gaussian mean value predicted by $\theta$.

\noindent\textbf{Dual Diffusion processes.}
Different from traditional diffusion model, we introduce two forward diffusion processes: positive diffusion process and negative diffusion process,  as shown in Fig.~\ref{framework}.
Specifically,
for the latent $z_{t-1}^{+}$ at timestep $t-1$ in the positive diffusion process, we add noise $\epsilon$ and obtain latent $z_{t}^{+}$. In a similar way, for the latent $z_{t-1}^{-}$ at step $t-1$ in the negative diffusion process, we add noise $-\epsilon$ and obtain latent $z_{t}^{-}$. Ultimately, we have,
\vspace{-0.05in}
\begin{equation}
\begin{aligned}
&q(z_t^{*}|z_{t-1}^{*})= \mathcal{N}(z_t^{*};\sqrt{1-\beta_{t}}z_{t-1}^{*},\beta_{t}I^{*}),\\
&q(z_{1:T}^{*}|z_{0}^{*})= \prod_{t=1}^{T}q(z_t^{*}|z_{t-1}^{*}),
\end{aligned}
\vspace{-0.05in}
\end{equation}
where $*$ denotes $+$ or $-$. Notably, for those two diffusion processes, they have the same starting point $z_0$ ($z_0^{+}=z_0^{-}$) and $I=I^{+}=-I^{-}$.

\noindent\textbf{Dual Reverse processes.}
In our work, we resort to the positive and negative conditioning to train the parameters of U-Net.  
We devise positive reverse process and negative reverse process to cover the original data $z_0$ from a probability density $p(z_T^{+})$ and $p(z_T^{-})$, respectively.
At timestep $t$, the latent $z_t^{+}$ is denoised to $z_{t-1}^{+}$ and the predicted noise is $\epsilon_\theta^{+}(t)$.
Similarly,
the latent $z_t^{-}$ is denoised to $z_{t-1}^{-}$ and the predicted noise is $\epsilon_\theta^{-}(t)$. Then, we have,
\vspace{-0.05in}
\begin{equation}
\begin{aligned}
&p(z_{0:T}^{*})= p(z_T^{*})\prod_{t=1}^{T}p_\theta(z_{t-1}^{*}|z_{t}^{*}),
\end{aligned}
\vspace{-0.05in}
\end{equation}
where $*$ stands for $+$ or $-$.
In essence, the common autoencoder $\theta$ is optimized with dual reverse processes, enhancing the D2M generation performance.


To establish the temporal correlation and rhythmic consistency during the training of conditional diffusion model,
we not only focus on extracting rhythmic visual cues and motion information from the dance video played in the normal forward direction, but also concern the distinct rhythms and motion information conveyed by reverse-played counterparts.
In this way, 
the desirable and undesirable temporal attributes of the generated music can be characterized.
Accordingly, 
the effectiveness of U-Net can be enhanced by these two conditionings.
Inspired by such assumption, as shown in Fig.~\ref{unet-figure}, we adapt U-Net of stable diffusion (SD)~\cite{RombachBLEO22} for the dance video conditioned music generation task.

\subsection{U-Net of PN-Diffusion Model} \label{3.3}

 \begin{figure}[t]
	\centering
\setlength{\abovecaptionskip}{0.05in}
	\includegraphics[scale=0.33]{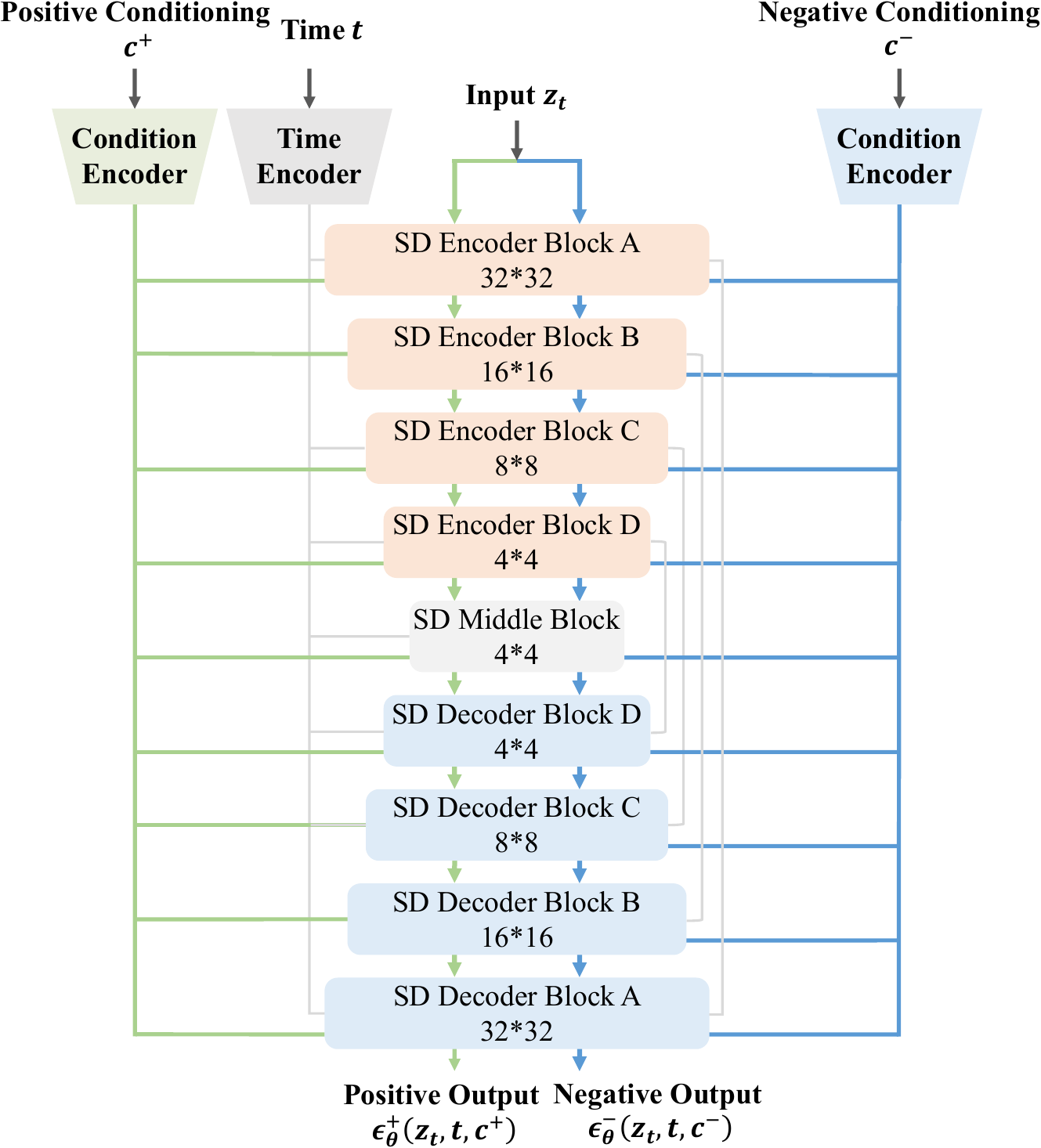}
\caption{Overview of the U-Net structure of PN-Diffusion model, where we adapt the U-Net of stable diffusion and use both positive and negative conditioning as input in the dual reverse process.
}\label{unet-figure}
 \vspace{-0.25in} 
\end{figure}


As shown in Fig.~\ref{unet-figure},
there are four-scales of stable diffusion encoder blocks and stable diffusion decoder blocks with one stable diffusion middle block.
Besides, both positive conditioning $c^{+}$ and negative conditioning $c^{-}$ are input to the U-Net, respectively.
Accordingly, we have the positive output $\epsilon_{\theta}^{+}(z_t,t,c^{+})$ and 
 $\epsilon_{\theta}^{-}(z_t,t,c^{-})$.
 To reverse the diffusion process to construct desired data samples from the noise,
 based on aforementioned positive conditioning and negative conditioning, we can train the conditional LDM via following Bi-directional denoising objective function,
\vspace{-0.05in}
 \begin{equation}
\begin{aligned}
&\mathbf{L}_{\epsilon}=\mathbb{E}_{\epsilon \sim N(0,I), z_t,t} 
\alpha \left[
\begin{Vmatrix}\epsilon-\epsilon_{\theta}^{+}(z_t^{+},t,c^{+})\end{Vmatrix}_2^2 \right]\\
&+(1-\alpha)\left[\begin{Vmatrix}-\epsilon-\epsilon_{\theta}^{-}(z_t^{-},t,c^{-})\end{Vmatrix}_2^2\right],
\end{aligned}
\vspace{-0.05in}
\label{loss}
\end{equation}
where $\alpha\in\left[0,1\right]$ is the nonnegative tradeoff parameters
to balance the contributions of two components in the loss function, allowing to learn the target model more effectively.
Besides, it is worth noting that bi-directional guidance is used to train the U-Net and predict $\epsilon$ in the training phase. In the inference phase, only the positive conditioning extracted from the normal forward-played dance videos is utilized to synthesis high quality accompaniment music.

%


\section{Experiment}

\subsection{Datasets}
In this work, we utilized two datasets with paired music and dance videos for the D2M evaluation: 
AIST++~\cite{LiYRK21} and TikTok~\cite{ZhuOWACYT22}.
In the training and testing phases, for fairness, we followed the standard evaluation scheme and dataset partition
of baseline method CDCD~\cite{Zhu0ORT023}, where
we split two datasets into training
and testing sets and only reported the performance on the testing set.

\noindent \textbf{AIST++.} 
AIST++ dataset was originally build from 
dance video database 
AIST~\cite{TsuchidaFHG19}.
AIST++ contains $1,020$ dance videos covering $10$ dance genres and corresponding music style with known camera poses, where each video is filmed in professional studios with a clean background.
Besides, it consists of $10$ types of music (\textit{i.e.}, lock, pop, and breaking) with $6$ songs for each type.
Notably, in our experiments,
the number of videos in training, testing, validation sets is $980$, $20$, and $20$, respectively. 
Moreover, for the baseline methods D2M-GAN~\cite{ZhuOWACYT22} and CDCD~\cite{Zhu0ORT023}, they both adopt $2$-second video and audio segments for training and testing in the main experiments.
Different from them, in our experiments, we manually split each long video into $5$-second video and audio segments for training and standard testing.
Finally, 
we thus obtained $20,140$ training and $234$ testing instances.

\noindent \textbf{TikTok.} 
TikTok Dance-Music dataset~\cite{ZhuOWACYT22} was originally collected from short video platform Tiktok, where $445$ dance videos with $85$ different songs are included. 
Following D2M-GAN and CDCD, we took $392$ videos to form the training set and kept the remaining $53$ videos as the testing set. 
Similarly,
we manually split each long video into $5$-second video and audio segments for training and standard testing.
Ultimately, we obtained $775$ training instances and $103$ testing instances.

In our experiment, 
for the visual frames of dance videos, we extracted RGB and flow features of visual frames using I3D model~\cite{CarreiraZ17} pre-trained on 
Imagenet~\cite{deng2009imagenet} and Kinetics~\cite{kay2017kinetics} datasets.
For the human body motions of dance performers, we employed BlazePose~\cite{abs-2206-11678} to obtain
2D skeletons, where the location of all 33 person keypoints are predicted for a single person from a single frame.

\subsection{Baselines}
To justify the effectiveness of our proposed PN-Diffusion, we chose five state-of-the-art recently proposed methods as baselines, including 
FoleyMusic~\cite{GanHCTT20}, D2M-GAN~\cite{ZhuOWACYT22}, CMT~\cite{DiJ0WZHLY21}, CDCD~\cite{Zhu0ORT023}, and LORIS~\cite{YuWCS023}.
 Note that we select
these baselines as their generation tasks are most relevant to our task, and their official codebases, involved parameters, and pre-trained models are all released.
In particular, D2M-GAN, CDCD, and LORIS are three newest methods for D2M generation task.
For FoleyMusic, it specifically deals with body-pose to rhythmic music generation.
For CMT, it addresses the task of video background music generation by establishing the rhythmic relations.
Above all,
we re-generated music segments for dance videos in our testing set and reported their music generation performance under their original experimental setting.

\subsection{Evaluation Protocols}

\noindent\textbf{Objective Evaluation.}
1) We follow the general paradigm of previous work~\cite{YuWCS023}
to measure 
the alignment of musical rhythms and dancing patterns.
The main metrics include improved versions of \textbf{beats coverage scores (BCS)} and \textbf{beats hit score (BHS)}.
Specifically, we assume that the rhythm point number of ground-truth music and generated music are $B_t$ and $B_g$, respectively.
Meanwhile, the aligned rhythm points are $B_a$.
Hence, BCS is calculated as the fraction of aligned rhythm points by the total beats from the generated music 
($B_a$/$B_g$), and BHS measures
the ratio of aligned beats to the ground truth beats
($B_a$/$B_t$).
Notably, we also 
modify the parameters of audio onset detection algorithms~\cite{BockKS12} 
 to avoid sparse rhythm vectors like LORIS.
In this way, 
BCS and BHS are more likely to play the roles of 
precision and recall, respectively.
Moreover, 
the F1 scores and the standard deviations of BCS and BHS (termed \textbf{CSD} and \textbf{HSD}) are measured to evaluate the generative stability. More details regarding these metrics can be found in LORIS~\cite{YuWCS023}.
2) Inspired by \textbf{Fr\'echet image distance(FID)} for image evaluation
and \textbf{Fr\'echet video distance (FVD)} for video evaluation, 
different from baseline methods,
we also adopt the \textbf{Fr\'echet audio distance (FAD)}~\cite{kilgour2018fr} to indicate the similarity between generated music and the ground-truth music.
In particular,
we utilize  three audio feature extractors: a) VGGish~\cite{HersheyCEGJMPPS17}, an audio classification model
trained on a large
dataset of YouTube videos.
b) PANNs~\cite{KongCIWWP20},
an audio neural networks trained on the large-scale AudioSet dataset. 
c) CLAP~\cite{wu2023large}, a contrastive language-audio model trained on the large-scale LAION-Audio-630K dataset.
And we name them as \textbf{FAD\_$v$}, \textbf{FAD\_$p$} and \textbf{FAD\_$c$}, respectively.

\noindent\textbf{Subjective Evaluation.}
In order to further make a comprehensive comparison to the competitive approaches, we conduct user study
to measure both the quality and the
relevance to the dance video of generated music.
Specifically, 
we invited volunteers 
 to perform a questionnaire survey for 
 the generated music samples of the AIST++ dataset.
 On the one hand, similar with AudioLDM~\cite{liu2023audioldm},
in each inquiry, we asked volunteers to
assign scores ranging from $1$ (bad) to $5$ (good) to evaluate the overall quality of the generated music sample (\textbf{OVL}) and 
the relevance of the generated music to the input conditional dance video (\textbf{REL}). 
We average these scores as the final score, namely Mean Opinion Score (MOS).
The higher the MOS score, the better music quality and the relevance between generated music and related dance video.
On the other hand,
similar with MM-Diffusion~\cite{ruan2023mm}, we
perform Turing Test for 
music samples generated by our
model and the ground-truth data. In details, 
we first blend them and then ask
volunteers to determine
whether they are generated and give the percentage that they are considered to be selected from the original AIST++ dataset.
\vspace{-0.05in}
\subsection{Implementation Details}
\noindent\textbf{Diffusion Model.}
In our work, the music is transformed to Mel-spectrogram and represented as image.
In the first stage, we train a perceptual compression models consisting of encoder $\mathcal{E}$
and decoder $\mathcal{D}$.
In this way, we can train the latent diffusion model in an efficient, low-dimensional
spectrogram latent space.
Specifically,
a DDPM is trained on a set of mel spectrograms, which can be used to synthesize similar mel spectrograms that are then converted back into music.
Following~\cite{RombachBLEO22},
we adopt linear noise schedule and noise prediction objective for all experiments, where
a conditional 2D U-Net model consisting of 4 scales of down blocks and up blocks is trained.
The diffusion step is set as $1,000$, and the whole model contains $166.55$M parameters.

\noindent\textbf{Parameter Setting.}
The sampling rate for all audio signals is $22,050$ $\mathrm{Hz}$ in our experiments. We use 5-second music samples for the main experiments, and
the resolution of the Mel-spectrogram is $256$.
For the first stage pretraining, 
when perform the Short-Time Fourier Transform (STFT) in audio processing, the stride between two adjacent time windows is set as $512$, and the window size of the Discrete Fourier Transform (DFT) is $2,048$.
The grid search strategy is adopted to determine the optimal values for the parameter $\alpha$. In addition, we
empirically set the batch-size to be $32$ and the maximum number
of iterations as $100$. The steps in the inference phase is set as $1,000$.
Besides, the FLOPs and parameters of our trained model are 
$905.77$G and $16.63$M. It takes about $12$ minutes to train one epoch on AIST++ dataset using 1 NVIDIA RTX A6000 GPU with $1000$ timestep.
\vspace{-0.05in} 

\subsection{Model Comparison}
To comprehensively evaluate the cross-modality correspondence between the dance video and generated music,
we report the 
quantitative results in terms of eight evaluation metrics in 
\cref{aist_tiktok}, and the best results are highlighted in bold.
 From these two tables, we can draw the following observations: 1)
Our PN-Diffusion consistently outperforms all the other baselines on AIST++ and TikTok datasets regarding BCS, CSD, BHS, HSD.
 In particular, with the best baseline LORIS on AIST++ dataset, PN-Diffusion 
achieves the significant improvement of $1.80$ and $4.22$ on BCS and BHS, respectively.
Besides, as for the TikTok dataset, our PN-Diffusion also 
has significant gains of $3.85$ and $5.90$ over baseline methods.
 This can be attributed to the fact that 
 both the positive and negative conditions are considered during the training of U-Net, making the parameters of the U-Net structure not only preserve the recovery
capability of the added positive noise in the positive diffusion process
but also retain the discriminative capability to predict the added negative noise in the negative diffusion process.
In this way, the ability of predicting
the positive noise can be enhanced benefiting from the inclusion of negative noise for comparison.
2) As for the F1 score of BCS and BHS, PN-Diffusion
achieves the impressive improvement of $2.01$ and $4.85$ on two datasets, indicating that the generative stability of PN-Diffusion is better than all baseline methods.
3)
Regarding the FAD$\_v$, FAD$\_p$ and FAD$\_c$,
except for the FAD$\_c$ on TikTok that we have a comparable result with the CDCD,
the performance of PN-Diffusion is significantly
better than all baselines and the numerical results are largely smaller than the baseline methods.
This confirms the fact that PN-Diffusion can enhance the
quality of generated music.

 \begin{table}[!t]
\centering
\setlength{\tabcolsep}{0mm}{
\scalebox{0.66}{
\begin{tabular}{
p{2.5cm}<{\centering}|
p{1.2cm}<{\centering}
p{1.2cm}<{\centering}
p{1.2cm}<{\centering}
p{1.2cm}<{\centering}
p{1.2cm}<{\centering}|
p{1.2cm}<{\centering}
p{1.2cm}<{\centering}
p{1.2cm}<{\centering}
}
\toprule
Method& BCS$\uparrow$ & CSD$\downarrow$  & BHS$\uparrow$ & HSD$\downarrow$&
F1$\uparrow$  & FAD\_$v$$\downarrow$ & FAD\_$p$$\downarrow$  & FAD\_$c$$\downarrow$\\
\midrule
\rowcolor{mgray}
\multicolumn{9}{c}{{AIST++}}
\\
\midrule
FoleyMusic~\cite{GanHCTT20}&  $92.00$ &  $13.33$ & $85.63$ &  $18.87$&$88.70$
&  $8.01$ &  $19.50$ & $1.10$ \\
D2M-GAN~\cite{ZhuOWACYT22}& $88.67$ & $10.49$  & $82.73$ & $16.86$& $85.60$& $11.29$& $27.76$& $1.48$
\\
CMT~\cite{DiJ0WZHLY21}& $\underline{95.92}$ & $8.19$  & $61.70$& $24.66$& $75.41$& $12.57$& $\underline{13.24}$& $1.02$
\\
CDCD~\cite{Zhu0ORT023}&  $92.18$  & $14.66$ & $80.50$& $21.16$&$85.95$ & $\underline{7.47}$& $18.06$& $1.25$
\\
LORIS~\cite{YuWCS023}& $95.84$ & $\underline{7.89}$  & $\underline{95.09}$ & $\underline{16.09}$& $\underline{96.45}$& $7.71$& $50.27$& $\underline{0.77}$
\\
\textbf{Ours}& $\mathbf{97.64}$ & $\mathbf{5.85}$  & $\mathbf{99.31}$ & $\mathbf{4.48}$& $\mathbf{98.46}$& $\mathbf{6.32}$& $\mathbf{4.35}$& $\mathbf{0.65}$
\\
\midrule
\rowcolor{mgray}
\multicolumn{9}{c}{{TikTok}}
\\
\midrule
D2M-GAN~\cite{ZhuOWACYT22}& $83.22$ & $30.03$  & $80.45$ & $30.66$& $81.81$& $27.30$& $13.26$& $1.46$
\\
CMT~\cite{DiJ0WZHLY21}& $85.42$ & $32.56$  & $60.03$ & $31.07$& $70.52$& $\underline{20.45}$& $15.56$& $1.30$
\\
CDCD~\cite{Zhu0ORT023}& $\underline{85.66}$ & $\underline{27.23}$  & $\underline{85.83}$ & $\underline{27.17}$& $\underline{85.75}$& $26.53$& $\underline{3.07}$& $\mathbf{1.11}$
\\
\textbf{Ours}& $\mathbf{89.51} $& $\mathbf{17.11} $  & $\mathbf{91.73}$ & $\mathbf{13.33}$& $\mathbf{90.60}$& $\mathbf{16.37}$& $\mathbf{1.14}$& $\underline{1.25}$
\\
\bottomrule
\end{tabular}
}
}
\vspace{-0.12in} 
\caption{
The quantitative comparison between PN-Diffusion and baseline music generation models conducted on AIST++ testing set and TikTok  testing set. The bold and underline indicate the best performance and the second best performance, respectively.
} 
\label{aist_tiktok}
\vspace{-0.25in} 
\end{table}

\begin{figure}[!t]
 \begin{subfigure}{0.49\linewidth}
 \centering
        \includegraphics[scale=0.045]{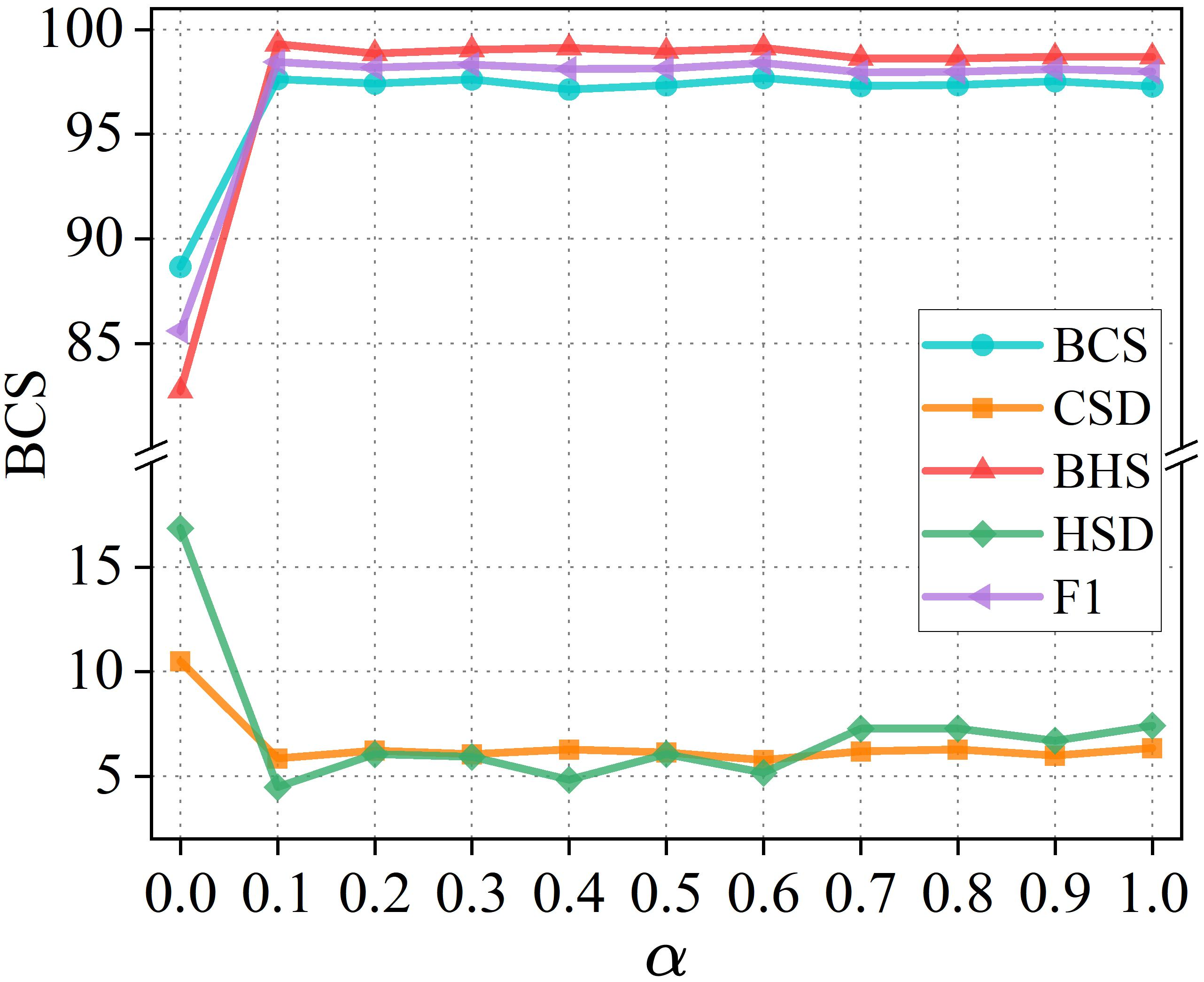}
  \caption{AIST++}
 \end{subfigure}
 \begin{subfigure}{0.49\linewidth}
 \centering
        \includegraphics[scale=0.045]{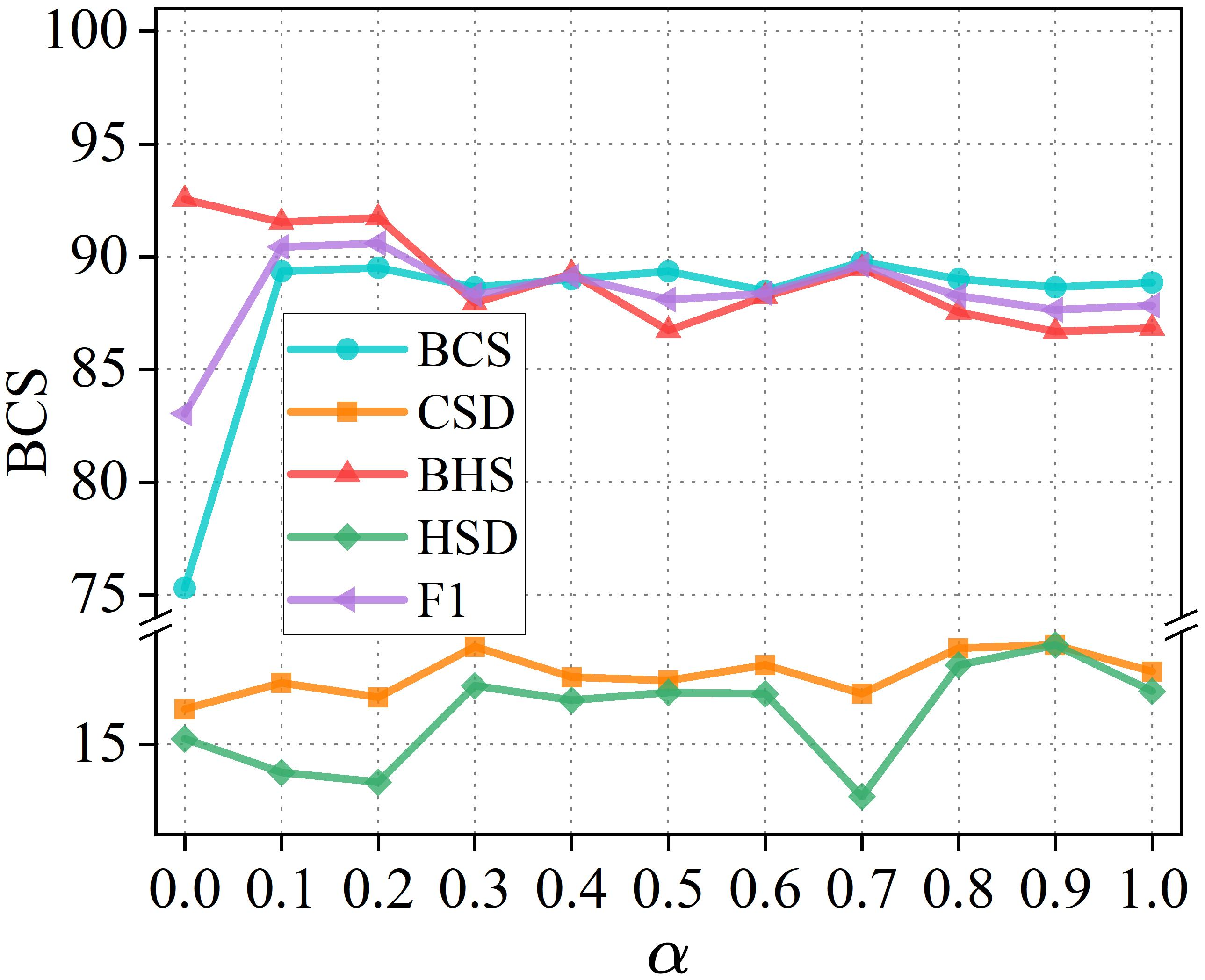}
  \caption{TikTok}
 \end{subfigure}
\vspace{-0.15in}
 \caption{Sensitivity analysis of the hyper-parameters $\alpha$.}
 \label{alpha}
 \vspace{-0.29in}
\end{figure}

\subsection{Component Analysis}
In our method, we have positive reverse process and negative reverse process corresponding to the positive conditioning and negative conditioning. Accordingly, there are two components in our final loss function in Eqn.~\ref{loss}.
To verify the effectiveness of each key component in our model,
namely, positive noise prediction and negative noise prediction, we investigate the nonnegative trade-off
parameter $\alpha$ in Eqn.~\ref{loss}.
 The sensitivity analysis of $\alpha$ on two datasets is
shown in Fig.~\ref{alpha}, where we vary $\alpha$ from $0$ to $1$ with a step of
$0.1$. 
As can be seen, the overall optimal performance can be achieved when $\alpha=0.1$ for AIST++ and $\alpha=0.2$ for TikTok,
indicating that both components are essential to train the diffusion model and have different contributions.

\subsection{Ablation Study}
To gain more deep insights, we further investigate the performance of the proposed PN-Diffusion when only the visual rhythm conditioning or the motion information conditioning is utilized as the conditioning, namely PN-V and PN-M. 
Here, we conduct experiments on the AIST++ and TikTok datasets, and the $\alpha$ is set as $0.1$ and $0.2$, respectively.
Besides,
to better explain the benefit of incorporating the negative diffusion process and negative reverse process to the LDM, we set two derivatives of our model, termed as P-Diffusion and N-Diffusion, where only positive conditioning is considered in P-Diffusion and only negative conditioning is considered in N-Diffusion. 
In our work, we found that the synchronization is not satisfactory when combining original music and reverse-played dance video together.
Thus, we resort to the synchronization between the original music and reverse-played dance video, and extract the positive and negative conditioning. 
To verify such observation and determine whether backward-playing is indeed comparatively optimal choice, we also tried another two intuitive methods.
First, we randomly selected video whose music is different from original dance music to mine negative conditioning, namely RN-Diffusion.
Second, the extracted positive features are directly negated to serve as the negative conditioning, namely DN-Diffusion.
The comparative
experimental results for these six variants of PN-Diffusion can be found in Tab.~\ref{ablation}.
As can be seen, overall, 
our PN-Diffusion consistently shows superiority over above four derivatives of PN-Diffusion, which validates the necessary of combining the visual rhythm cues and motion information together (compare with PN-M and PN-V) as well as the importance of considering the positive and negative conditioning (compare with P-Diffusion and N-Diffusion) to train the D2M latent diffusion model.
Meanwhile, the results compared with RN-Diffusion and DN-Diffusion validate the effectiveness of choosing the reverse-played dance videos to extract negative conditioning.
One plausible explanation is that 
reverse-played videos in PN-Diffusion retain the same poses, transitions, and temporal structure as forward-play videos but in the opposite direction, creating a more faithful ``negative" pairing.
In contrast, RN-Diffusion
introduces different body movements, camera angles, and overall style from randomly selected video, while DN-Diffusion does not incorporate any real temporal inversion.

\begin{table}[!t]
\centering
\setlength{\tabcolsep}{0mm}{
\scalebox{0.63}{
\begin{tabular}{p{2cm}<{\centering}|p{1.1cm}<{\centering}p{1.1cm}<{\centering}p{1.1cm}<{\centering}p{1.1cm}<{\centering}
p{1.1cm}<{\centering}|p{1.1cm}<{\centering}p{1.1cm}<{\centering}p{1.1cm}<{\centering}p{1.1cm}<{\centering}
p{1.1cm}<{\centering}
}
\toprule
\multirow{2}[1]{*}{Method} & \multicolumn{5}{c|}{AIST++}&\multicolumn{5}{c}{TikTok}\\
\cline{2-11} 
&BCS$\uparrow$ & CSD$\downarrow$  & BHS$\uparrow$ & HSD$\downarrow$&
F1$\uparrow$& BCS$\uparrow$ & CSD$\downarrow$  & BHS$\uparrow$ & HSD$\downarrow$&
F1$\uparrow$\\
\hline
PN-M&  $94.48$ &  $5.34$ & $88.88$ &  $14.47$&$91.59$& $78.23$ &  $18.04$ & $90.80$ &  $14.99$&$84.05$\\
PN-V&  $96.43$ &  $6.37$ & $98.20$ &  $5.95$&$97.31$&   $86.74$ &  $17.89$ & $85.85$ &  $16.76$&$86.30$\\
\hline
P-Diffusion&  $96.33$ &  $5.50$ & $96.83$ &  $6.70$&$96.58$&  $78.39$ &  $18.09$ & $76.34$ &  $17.23$&$77.35$\\
N-Diffusion&  $85.80$ &  $10.51$ & $80.51$ &  $17.82$&$83.07$&  $67.99$ &  $18.06$ & $80.97$ &  $14.04$&$73.91$\\
\hline
RN-Diffusion& $88.73$ &  $10.60$ & $83.84$ &  $18.17$&$86.22$
&  $85.63$ &  $18.65$ & $85.74$ &  $16.25$&$85.68$\\
DN-Diffusion& $93.43$ &  $6.29$ & $94.61$ &  $8.29$&$94.02$
&  $83.85$ &  $17.26$ & $81.47$ &  $18.19$&$82.64$\\
\hline
Ours& $\mathbf{97.64}$ & $\mathbf{5.85}$  & $\mathbf{99.31}$ & $\mathbf{4.48}$& $\mathbf{98.46}$& $\mathbf{89.51}$ & $\mathbf{17.11}$  & $\mathbf{91.73}$ & $\mathbf{13.33}$& $\mathbf{90.60}$\\
\bottomrule
\end{tabular}
}}
\vspace{-0.1in}
\caption{Ablation studies on the conditioning type.} 
    \label{ablation}
\vspace{-0.3in}
\end{table}

\subsection{User Study}
To more comprehensively evaluate the quality of generated music and the cross-modal correlation between generated music and the condition dance video in terms of dance-music beats, we provide complementary
qualitative demos of different methods, where the dance segments corresponding to different times are similar.
We also provide some generated music exmaples of CDCD\footnote{\url{https://youtu.be/_7VHPpKLlSM.}}, LORIS\footnote{\url{https://youtu.be/KUdGDD2MqR4.}}, and PN-Diffusion\footnote{\url{https://youtu.be/J0YNWEWRTHY.}}.

By comparing these dance videos, we can easily draw the conclusion that the performance
of PN-Diffusion in generating highly-synchronized and high-quality music
for the given dance video is better than baseline methods.
In details, 
for both CDCD and LORIS, they mainly produce music that is merely a collection of rhythmic sounds and is significantly different from the actual music needed.
In contrast, the music generated by our method is more in line with real-world requirements.

Besides,
we also conduct user study to compare our PN-Diffusion with CDCD and LORIS on the AIST++ dataset.
Specifically, 
we invited
 $10$ volunteers who are research scholars actively 
engaging in cutting-edge computer vision fields and are also music 
lovers
to perform a questionnaire survey for 
 the generated music samples of the AIST++ dataset, where $10$ generated music samples are randomly selected 
 \begin{wraptable}{r}{0.29\textwidth}
\vspace{-0.15in} 
\centering
   \scalebox{0.7}{
\begin{tabular}{p{2.4cm}<{\centering}|p{1.2cm}<{\centering}p{1.2cm}<{\centering}
}
\toprule
\multirow{2}{*}{Method} & \multicolumn{2}{c}{AIST++} \\
\cline{2-3}
 & OVL$\uparrow$ & REL$\uparrow$\\
    \hline
CDCD~\cite{Zhu0ORT023}&  $2.2$ &  $2.1$ \\
LORIS~\cite{YuWCS023}& $2.0$ &  $2.0$ \\
\textbf{Ours}&  $\mathbf{4.3}$ & $\mathbf{4.2}$ \\
\bottomrule
\end{tabular}
} 
 \vspace{-0.1in} 
\caption{Mean opinion score on AIST++.}\label{user}
\vspace{-0.15in} 
\end{wraptable} and report the subjective results.
 for each tester.
As can been seen from Tab.~\ref{user}, compared the CDCD and LORIS, 
testers found that the music samples generated by our method are better, both in the quality of the music and its synchronization with dance videos. 
Regarding the music quality, 
volunteer perceived that our generated musics are more akin to real-world music, rather than just rhythmic sounds.
Besides, we randomly selected $10$ generated music for each tester and conducted the Turing Test for the generated music.
Specifically, the probability of passing the Turing Test where the ground-truth music can be identified by tester is $93.9\%$, 
Meanwhile, the percentage that our generated music can successfully cheat the tester and mistakenly be regarded as the ground-truth music of AIST++ is $73.4\%$, proving the high quality and realism of music generated by our model once again.

\begin{figure}[!t]
 \begin{subfigure}{0.49\linewidth}
 \centering
  \includegraphics[scale=0.15]{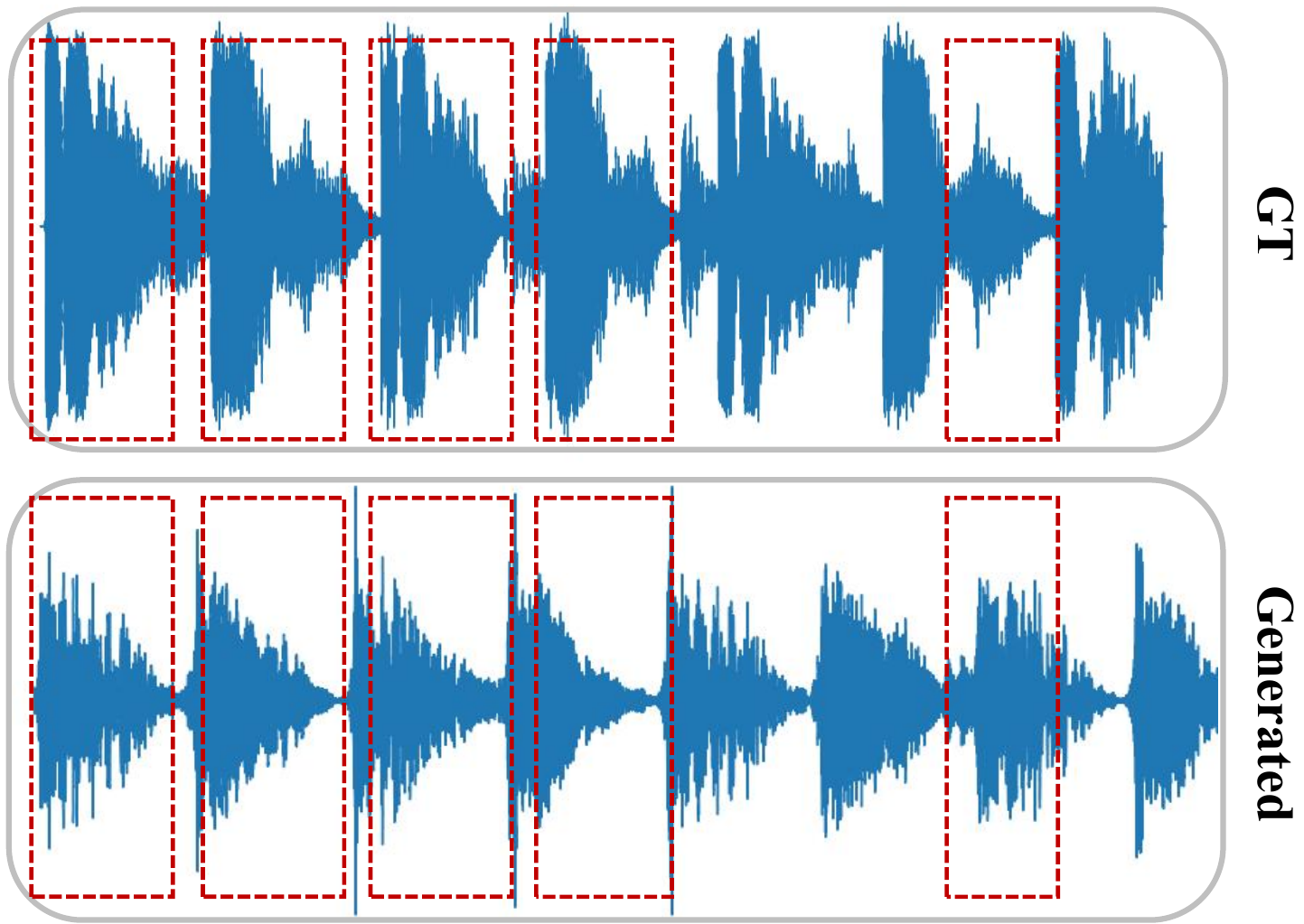}
  \caption{PN-Diffusion(Ours)}
  \label{our_wav}
 \end{subfigure}
 \begin{subfigure}{0.49\linewidth}
 \centering
  \includegraphics[scale=0.15]{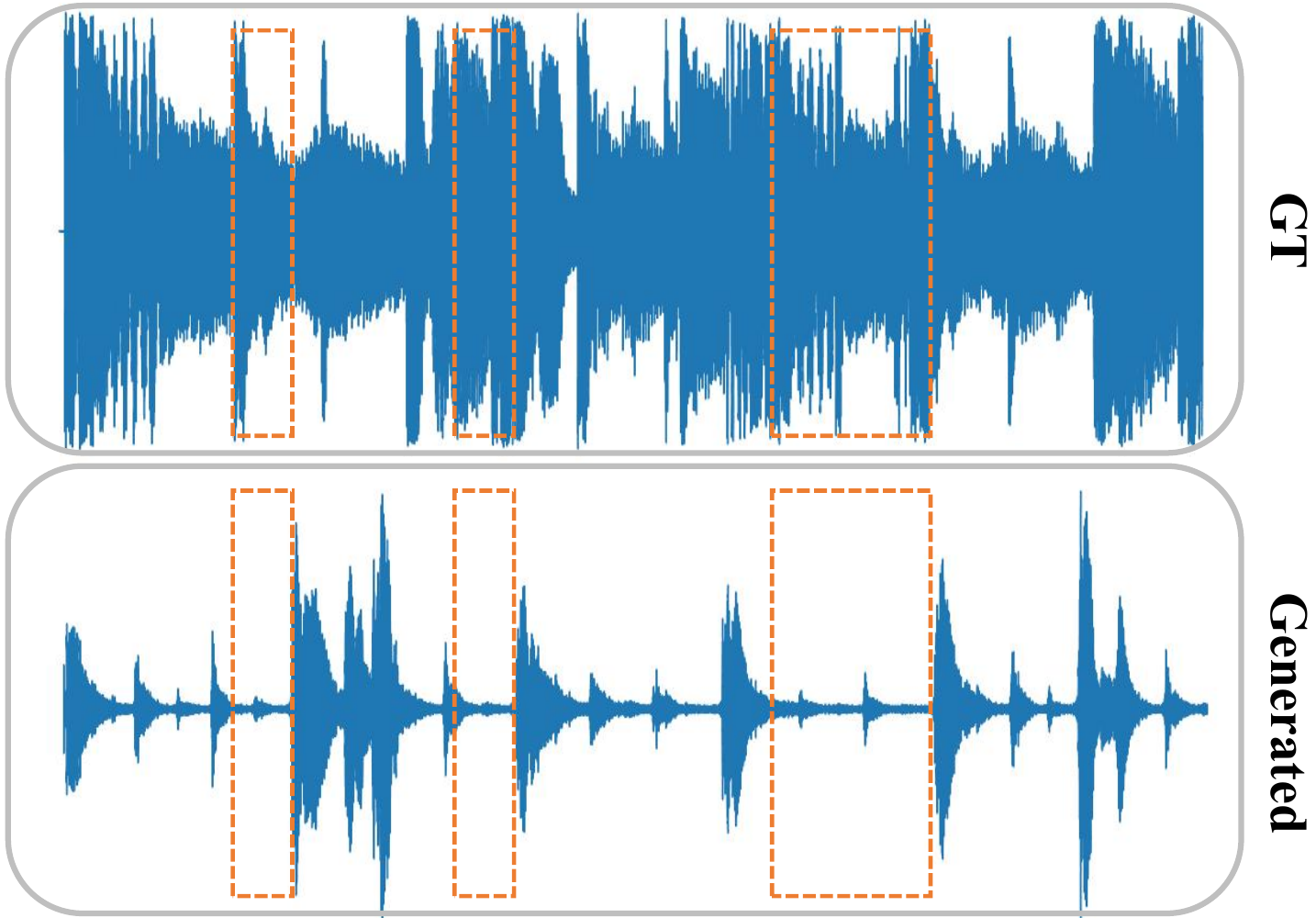}
  \caption{LORIS}
  \label{loris_wav}
 \end{subfigure}
\vspace{-0.15in}
 \caption{Waveform comparison of generated and GT music.}
 \label{waveform}
 \vspace{-0.3in}
\end{figure}

\subsection{Qualitative Results}
Apart from the quantitative analysis, we also provided certain
intuitive examples to illustrate the effects of our scheme~\cite{song2018neural,song2017neurostylist,zheng2023egocentric,zheng2021collocation,wu2023flow,sun2024dct}.
Here we choose LORIS as our baseline since it achieves overall best performance compared to other baselines.
Although PN-Diffusion and LORIS share the same datasets, the video lengths they processed are differ, where 
PN-Diffusion divides each long video into 5-second segments, while LORIS splits them into 25-second and 50-second segments.
LORIS generates longer segments, providing more global context but potentially missing local synchronization details, such as rhythm and pitch in shorter time spans, and 
PN-Diffusion focuses on shorter segments, emphasizing finer synchronization details. 
Therefore,
directly comparing the same GT audio may bias the evaluation, failing to accurately reflect each method's strengths and limitations.
For fairness, we provide 5-second waveform of generated music and the ground-truth (GT) dancing music in Fig.~\ref{waveform}.
Intuitively, PN-Diffusion generates highly synchronized audio with original GT dancing music from the perspectives of amplitude, frequency, and duration.
To be more specific, in Fig.~\ref{our_wav}, we highlight the 
sound slots using red dotted boxes, where the GT dancing music and our generated music have consistent rhythm and similar pitch. Obviously, our generated music is highly-synchronized with the GT dancing music.
Besides, in Fig.~\ref{loris_wav}, the 
waveform segments that are inconsistent in both pitch and sound rhythm are highlighted using the orange dotted boxes.
The GT dancing music is very continuous and has a long-term high pitches, while the music generated by LORIS is intermittent and has relatively low pitch, indicating that the generated music is inconsistent in rhythm with the corresponding dance video.
Reflecting on the  dance videos, there are instances where dance movements are present, but there is no corresponding music.

\begin{figure}[!t]
 \begin{subfigure}{1.0\linewidth}
 \centering
  \includegraphics[scale=0.2]{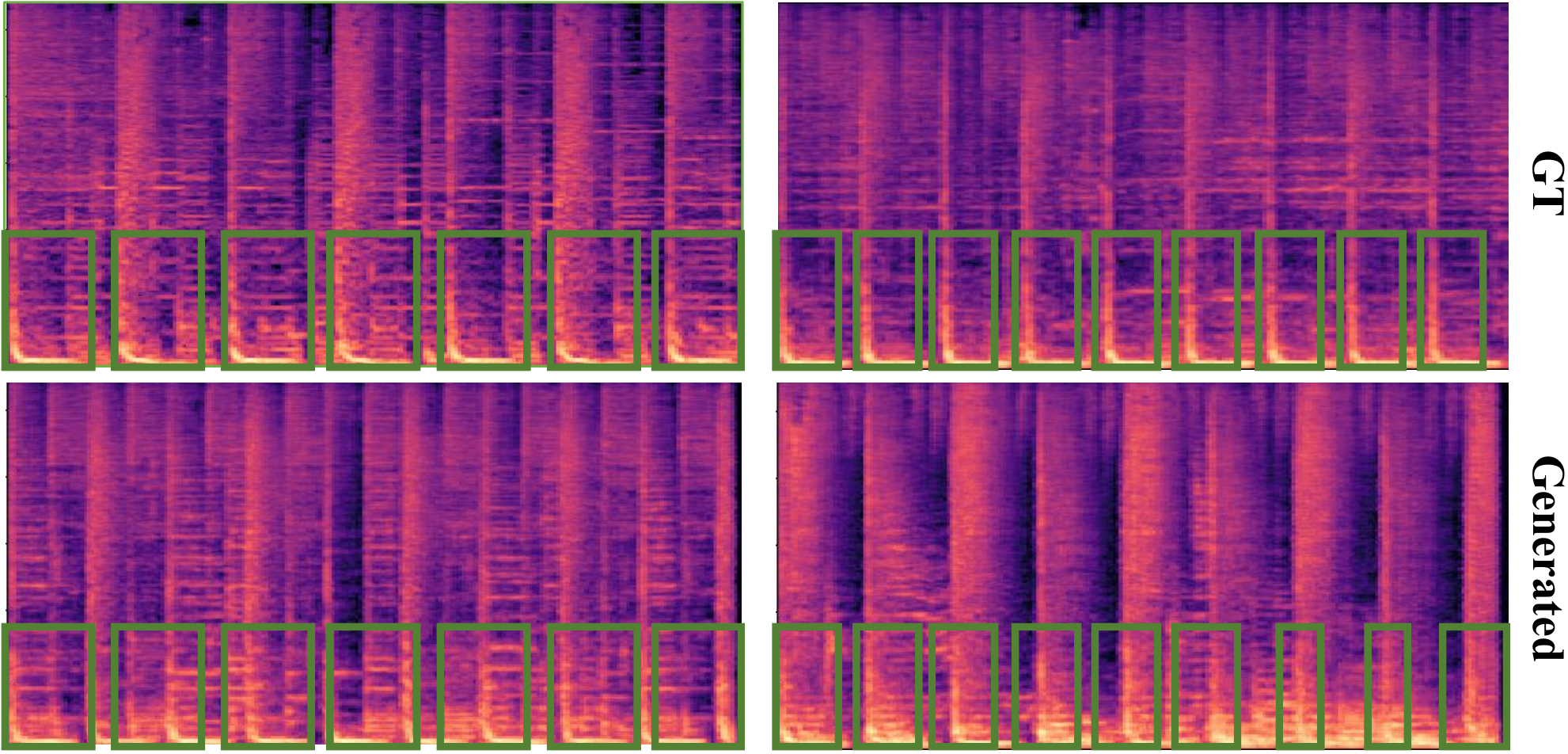}
  \caption{PN-Diffusion(Ours)}
  \label{our_mel}
 \end{subfigure}
 \\
 \begin{subfigure}{1.0\linewidth}
 \centering
  \includegraphics[scale=0.2]{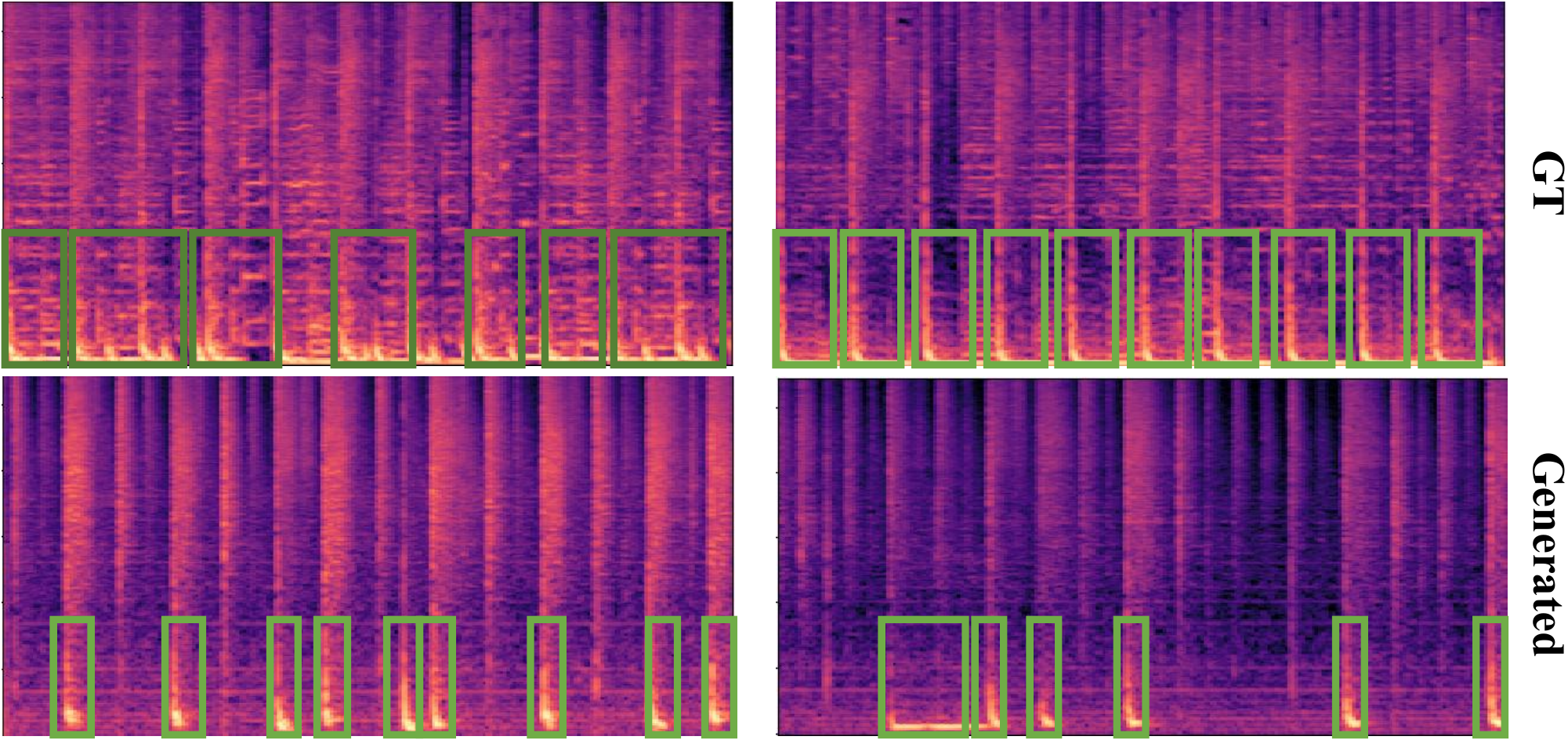}
  \caption{LORIS}
  \label{loris_mel}
 \end{subfigure}
 \vspace{-0.3in}
 \caption{Mel spectrogram Visualizations of GT dancing music and generated dancing music by PN-Diffusion and LORIS, where the areas with high intensity and energy are highlighted in green color boxes. Compared with LORIS, our generated music closely match the ground truth in terms of timing.
 }
 \label{mel}
 \vspace{-0.25in}
\end{figure}

Besides, to illustrate the spectrogram analogousness, we also visualize two pairs of Mel spectrogram correspond to the generated music and GT dancing music of PN-Diffusion and LORIS in Fig.~\ref{mel}, where the areas with high intensity and energy are highlighted in green color boxes.
As can be seen from Fig.~\ref{our_mel}, our generated music has a similar intensity and energy distribution and time alignment with the GT music.
In comparison, in Fig.~\ref{loris_mel}, 
there is a significant difference between the intensity and energy distributions of music generated by LORIS and GT music. Meanwhile, the shapes and intensities of frequency peaks across corresponding regions in both spectrograms of Fig.~\ref{loris_mel} are particularly dissimilar.
Above all, the superior performance of PN-Diffusion in  generating highly-synchronized music for the given dance video is further verified.


\section{Conclusions}
We focus on enhancing the dance to music synthesis via negative conditioning latent diffusion model, where positive conditioning and negative conditioning serve as bi-directional guidance to train the U-Net network.
Specifically, we introduce dual diffusion process and reserve process to fit 
 positive and negative rhythmic visual cues and motion information captured by playing dance videos forward and backward, respectively.
 Superior performances are achieved over two widely-used dance video datasets through objective and subjective evaluations, which can be attributed
to the new formulation of noise prediction loss function.

\noindent \textbf{Acknowledgments:} This research is supported by NSF IIS-2525840, ECCS-2123521 and Cisco Research unrestricted gift. This article solely reflects the opinions and conclusions of its authors and not the funding agencies.

{
    \small
        \bibliographystyle{unsrt}%
    \bibliography{main}
}


\end{document}